\documentclass{elsart}

\newcommand{\m}{\mathbf}

\usepackage{natbib}

\usepackage{amssymb}
\usepackage{amsmath}

\begin{document}

\begin{frontmatter}



\title{Chiral scattering in complex magnets}


\author{S. V. Maleyev}

\address{Petersburg Nuclear Physics Institute, Gatchina, Leningrad District 188300, Russia}

\ead{maleyev@sm8283.spb.edu}

\begin{abstract}
% Text of abstract
General properties of the chiral scattering of polarized neutrons are considered for two possible axial vector interactions: Zeeman energy and non-alternating Dzyaloshinskii-Moriya interaction. Behavior in magnetic field  of helical magnetic structures is discussed for $Mn Si$ and magneto-electric materials. The dynamical chiral fluctuations in  magnetic field are considered briefly. The chiral fluctuations  in materials with the Dzyaloshinskii-Moriya interaction are discussed and an assumption is made that above the transition temperature they have to be incommensurate.
\end{abstract}


\end{frontmatter}

\section{Introduction}
Spin chirality (SC) is determined as a vector product 
$\m{S_R\times S_{R'}}$. It  enters in the right-hand side of the expression for the time derivative $d\m{S_R}/d t$ and  known also as spin current. Its study is very important for complete understanding of some magnetic systems.  For example the SC along with the spin density are relevant variables for the second order  phase transition in frustrated triangular-lattice antiferromagnets \cite{K}. This theoretical suggestion  was experimentally confirmed in \cite{P1}, \cite{P2} (see also \cite{M1}, \cite{M2}).

There are two kinds of the spin chirality: i) Static chirality which displays itself as helical structure in ordered magnets. ii) The chiral fluctuations. However the  last are described by the four spin  correlation function $\langle \m{[S_R\times S_{R'}],[S_{R_1}\times S_{R'_1}]}\rangle $. Unfortunately it can not be  measured by existing methods.

However  using polarized neutrons one can study fluctuations in the chiral channel which are described by vector chirality  \cite{M1}, \cite{M2}
\begin{equation}
\begin{aligned}
\m C(\m Q,\omega)&=-\frac{1}{2}\int_0^\infty d
t e^{i\omega t}\\&\langle \m{S_{-Q}}(t)\times \m{S_Q}(0)+\m{S_Q} (0)\times \m{S_{-Q}}(t) \rangle ,
\end{aligned}	
\end{equation}
connected to the antisymmetric part of the magnetic susceptibility:  $\chi_{\alpha \beta}-\chi_{\beta \alpha})/2=-i\epsilon_{\alpha \beta \gamma}C_\gamma$.

The chiral contribution to the magnetic neutron scattering is given by \cite{M1},\cite{M2}
\begin{equation}
	\sigma_C(\m Q,\omega)=N_0 \frac{2(\m P\cdot\hat Q)(\hat Q\cdot I m \m C(\m Q,\omega))}{1-\exp(\omega/T)},
\end{equation}
where $\hat Q=\m Q/Q$, $\m P$ is the neutron polarization  and the factor $N_0$ is the same as in conventional magnetic scattering.

The scattering intensity  and  $\m P$ are  a scalar and an axial vector respectively. So $\m C$ has to be an axial vector also. It is possible if the system has an intrinsic axial vector too. There are following possibilities: i) Constant magnetic field or the sample magnetization. ii) Non-alternating Dzyaloshinskii-Moriya interaction. iii) Magnetic helical structure with definite sense of the spin rotation. We consider all these possibilities below.

\section{Static chirality}

We begin with simple ("ferromagnetic") helix \cite{M1}, \cite{M3} \cite{B} where for the lattice spin we have
\begin{equation}
	\m{S_R}=S[\hat z\sin\alpha+(A e^{i\m k\cdot \m R}+A^* e^{-i\m k\cdot\m R})\cos\alpha],
\end{equation}
	where $A=(\hat x-i\hat y)/2$ and $\hat x,\hat y, \hat z$ are  orthogonal unit vectors. If $\alpha=0$ and $\alpha=\pm\pi/2$ we have plain helix and ferromagnet respectively.
	
	Replacing $\hat z\to \hat z\cos\m k_{AF}\cdot\m R$ and $\m k\to \m k+\m k_{AF}$ we get antiferromaghetic helix which becomes an antiferromagnet  if $\alpha=\pm \pi/2$.
	
	For the simple helix the scattering intensity is given by
\begin{equation}
\begin{aligned}
	I&=I_o+I_-+I_+,\quad
	I_0=N_0\sin^2\alpha(1-\hat Q^2_z) \delta(\m Q+\m K),\\&
	I_\pm=N_0(\cos^2\alpha/4)\{1+\hat Q^2_z
	\pm 2(\m P\cdot \hat Q)(\hat Q\cdot [\hat x\times\hat y])\}\delta(\m Q\pm \m k+\m K),
	\end{aligned}
\end{equation}
where $\m K$ is the reciprocal lattice point. For the AF helix in these equations  we must replace $\m K\to \m K_{AF}$.

The sum $I_-+I_+$ changes sign if  $\m k\to-\m k$. In centrosymmetrical crystals the helical structure is degenerated in respect of sign of the vector $\m k$ and real sample consists of domains with opposite $\m k$ directions. In the case of equal domain population the chiral scattering disappears. Unequal  population may be accidental as was observed in $Cs Mn Br_3$ \cite{P2} or artificially prepared by cooling the sample below $T_N$ at an inversion symmetry breaking condition. In the case of $Zn Cr_2 Se_4$ the single domain state was attained by cooling in crossed electric and magnetic fields \cite{S}.
 In $Ho$ unequal domain population was obtained using elastic torsion \cite{P3}. In both these cases the external action works as the intrinsic Dzyaloshinskii-Moriya interaction which  fixes a single $\m k$ structure and leads to strong chiral scattering.
 
 To the best of my knowledge the non-centrosymmetric cubic  $Mn Si$ helimagnet was the only compound where the static chirality  was studied throughout \cite{Sh}, \cite{I}, \cite{G1}, \cite{G2}. In $Mn Si$ we have $\m k=(2\pi/a)[\xi,\xi,\xi]$ where $\xi\simeq 0.017$,  $a=0.4558nm$ and the spins rotate in the perpendicular plane \cite{Sh}, \cite{I}. In complete agreement with Eq.(4) the polarized neutrons allow to suppress one of the $\pm \m k$ satellites of the Bragg scattering \cite{I}.
 
  However in  the magnetic field  new features were found:
i) Ferromagnetic spin configuration takes place in the field $H_\parallel$ along the vector $\m k$ if $H_\parallel>H_{C\parallel}\simeq 0.6T $ \cite{Ko}. According \cite{M4} we have $H_{C\parallel}\simeq A k^2$ where $A$ is the spin-wave stiffness. This expression is  in agreement with existing experimental data. Below $H_{C\parallel}$ we have $\sin\alpha=-H_\parallel/H_{C\parallel}$. Obviously above $H_{C\parallel}$ the static chirality disappears but it has to remains in the spin-wave channel \cite{M4}. ii) In the low perpendicular field $H_\perp$ the second  harmonic $\m k_2=2\m k$ appears with corresponding chiral contribution. With increasing of $H_\perp$ the vector $\m k$  rotates toward the field and at $H\sim \Delta\sim 0.1T$ where $\Delta$ is the spin-wave gap it turns along the field. Simultaneously the second harmonic disappears \cite{G1}, \cite{G2}. This behavior was explained in \cite{M4} as a consequence of mixing spin-waves with momenta $\m q$ and $\m q\pm\m k$ in perpendicular field and   Bose condensation of the spin-waves with momenta $\m q=0$ and $\pm \m k$. Similar phenomena were observed in $Ba_2Cu Ge_2O_7$ [\cite{Z} and references therein]. However in these papers the chirality has not been studied.

We discuss in such details  the $Mn Si$ helical structure transformations in magnetic field having in mind new magneto-electric materials $R Mn O_3$ ($R=Gd,Tb,Dy$), $Ni_3V_2O_8$ etc. where helical magnetic structure and the DM interaction appear simultaneously with the inversion symmetry breaking transition \cite{Ka}, \cite{Ke}, \cite{Mo}, \cite{Se}, \cite{H}.

 As the static chirality  takes place in any helical magnets its experimental study in these magneto-electrics can elucidate real magnetic structures and their transformations in magnetic field differently directed relative the crystal axes. Preliminary theoretical consideration of the related problems will be published soon \cite{Sy}. 
 
 \section{Chiral fluctuations; general properties}     

Properties of the chiral fluctuations including  the spin-wave scattering are determined by the symmetry properties of the axial vector responsible for the chirality. This problem was considered
throughout in \cite{M1} for two cases i) Magnetic field and ii) Non-alternating DM interaction. We present here the final results only. 

i) {\it{Dynamical chirality}}. Magnetic field is $t$-odd: it changes sign at time reflection  $t\to -t$ and $I m\m C(\m Q,\omega)$ is even function both $\omega$ and $\m Q$. In classical limit when $\omega\ll T$  we can replace the Bose factor in Eq.(2) by $T/\omega$. As a result the chiral scattering changes sign with $\omega$ in agreement with experiment \cite{P1}, \cite{P2}, \cite{G} and has to be zero at $\omega=0$. So we have hot static chirality in magnetic field. In this approximation the $\omega$ integrated chiral scattering is zero. 

ii){\it{The DM chirality.}} The DM interaction between two spins has the form
\begin{equation}
	V_{DM}=\m D_{\m R_1 \m R_2}[\m S_1\times \m S_2]. 
\end{equation}
It can not change if $1\Leftrightarrow 2$. So we have 
\begin{equation}
	\m D_{\m R_1 \m R_2}=-\m D_{\m R_2 \m R_1},
\end{equation}
 or in momentum space $\m D_\m Q=-\m D_{-\m Q}$. Besides the Dzyaloshinskii vector is $t$-even. As a result $I m\m C(\m Q,\omega)$ is  the odd function of both $\m Q$ and $\omega$. Particular in the classical limit the $\omega$ integrated DM chirality is not zero but changes sign with $\m Q$.
 
 In noncentrosymmetric crystals in magnetic field the function $I m\m C$ has not simple symmetric properties. As an example  the spin-wave scattering in cubic magnets with the DM interaction in strong field was considered in \cite{M4}, and in antiferromagnets \cite{Sy1}.
 
 In the centrosymmetric crystals below $T_N$ the spin-wave chiral scattering have the same symmetry as in the DM case if one has the single domain sample. 
 
 \section{Dynamical chirality}
 
 We do not give here detailed survey of the of the dynamical chirality as in may be found in \cite{M1}, \cite{M2} and mention only recent achievements in this field.
 
 In \cite{L} the chiral scattering in spin-singlet quasi-1D compound $Sr_{14}Cu_{24}O_{41}$ was studied. The singlet triplet transitions with different chiral projections of the triplet spin were examined.
 
 Small-angle critical chiral scattering in ferromagnet $Eu S$ was studied in \cite{G5}. In weak magnetic field this scattering is described by the three spin correlation function with  the neutron and the field momenta $\m Q$ and zero respectively. According  theory (so called  Polyakov- Kadanoff-Wilson operator algebra \cite{Po}, \cite{Kad}, \cite{Wi}) if in multi-spin correlation function the momentum $Q>>$ inverse correlation length and other momenta the $Q$ dependence appears as a factor $Q^{-(5+1/\nu)}$ where $\nu$ is the correlation length exponent. As a result the chiral scattering becomes proportional to $(T-T_c)^{-\nu}$ (see \cite{M1} and references therein). For ferromagnets $\nu\simeq 0.67$. This nontrivial  theoretical prediction was checked for iron with $\nu=0.67(7)$. Corresponding result for  $Eu S$ is $\nu=0.64(5)$. So for the second order phase transition the operator algebra is now confirmed experimentally.

  \section{DM chirality}
 
  To the best of my knowledge up to now the DM chiral fluctuations  were studied in $Mn Si$ only \cite{R} \cite{G4}. I know also an  attempt to observe them in $Cs Co Cl_3$ \cite{P3}. In \cite{G4} the data of the small angle polarized neutron scattering above $T_N$ are mainly in agreement with simple mean field theory which takes into account the following interactions \cite{BJ}: conventional exchange, isotropic DM interaction and anisotropic exchange. Corresponding expression for the scattering intensity is given by \cite{G4}
\begin{equation}
	I(\m Q)\sim\frac{k^2+Q^2+\kappa^2-2k \m Q\cdot \m P}{(Q-k)^2+\kappa^2+(g k)^2([\hat Q^4]-1/3)},
\end{equation}
where $k$ is approximately equal to the helix wave-vector at low $T$ (see Sec. II), $[\hat Q^4]=\hat Q^4_x+\hat Q^4_y+\hat Q^4_z$ is a cubic invariant,  $g$ is a constant proportional to the strength of the anisotropic exchange and $\kappa$ is an inverse correlation length. Usually near the second order phase transition $\kappa\sim(T-T_N)^\nu$.
 In the mean field theory $\nu=1/2$. However it was found that $\nu=O.62(1)$. Moreover it was found that the last term in the denominator has to be replaced by $([\hat Q^4-1/3])^{0.22(5)}$. These new critical exponents demand theoretical explanations. 
 
 General form of the critical scattering is unusual also. The last $g^2$ term in the denominator is very small. Neglecting it  for $\m P=0$ we obtain that  the critical scattering is maximal at the sphere $Q=k$ instead of to be maximal at the positions of the Bragg peaks below $T_N$. For the fully polarized neutrons the chiral term in Eq.(7) suppresses the scattering at $\m Q=\m P$ and the scattering intensity has form of half-moons depending on sign of $\m P$ as was observed in \cite{G4}.
 The intensity is peaked at the Bragg positions        
very close to $T_N$ where $\kappa$ becomes comparable with $g k$. Expression $[\hat Q^4]-1/3$ has minima  equal to zero along cubic diagonals and if $\kappa\ll gk$ we have conventional critical fluctuations in these directions.

For $Mn Si$ we have incommensurate critical fluctuations. Apparently it is a general situation in the case of the non-alternating DM interaction. This statement is confirmed  by exact solution of the 1-D chain with the DM interaction and  3-D mean field calculations \cite{A} \cite{M1}.

In this respect the mean field results for the critical fluctuations in weak ferromagnets such as $Mn C O_3$ are very instructive. In this case there are three principal interactions: isotropic exchange, easy plain anisotropy and the DM interaction in perpendicular $z$ direction.
After simple calculations (cf. \cite{M1}, \cite{G4}, \cite{A} ) we have
\begin{equation}
	I=N_0\{G_z(\m Q)(1-\hat Q^2_z)+G_\perp(\m Q)[1+\hat Q^2_z-2\Lambda q_z(\m P\cdot \hat Q)] \}
	\end{equation} 
where $\Lambda$ is proportional to the strength of the DM interaction, vector $\m q$ is the distance to the nearest antiferromagnetic Bragg point  and the functions $G_z$ and $G_\perp$ are given by
\begin{equation}
	G_z(\m Q)=\frac{V}{T-T_z+[A\m q]^2},
	\end{equation}
\begin{equation}
G_\perp(\m Q)=\frac{V\{T-T_0+[A\m q]^2\} }{\{T-T_\perp+[A(\m q-\Lambda \hat z)]^2\} \{T-T_\perp+[A(\m q+\Lambda \hat z)]^2\}},
\end{equation}
where  $[A\m q]^2=A_\perp q_\perp^2+A_z q^2_z$,  $T_0$ is the transition temperature if one neglects both the anisotropy and DM interaction, $T_z=T_0-\beta$ where $\beta$ is the strength of the anisotropy and $T_\perp=T_0-\Lambda^2/(4A_z)$.

We see that there are two mean field transition temperatures and incommensurate fluctuations in the perpendicular and chiral channels. At the same time the low $T$ weak ferromagnets are  commensurate structures. So there are two possibilities: i) There is the first order transition to the weak ferromagnetic state; ii) the critical fluctuations suppress the incommensurability. The problem demands further experimental and theoretical investigation. 

\section{Conclusions}
The spin chirality is an important complementation to conventional magnetic neutron scattering if the system as whole has  an intrinsic axial vector such as the magnetic field and the non-alternating DM interaction.

 The static chirality (helical magnetic structure) is sensitive to external magnetic field and its study may be useful for understanding some features of new  magneto-electric materials.

The Dynamical chirality  has provided nontrivial information in different fields of magnetism and  became an usual tool of magnetic neutron scattering.

The study of the DM chiral fluctuations is now at very beginning but promises unusual results. 

\section{Acknowledgments} 
  This work is partly supported by RFBR (Grants $03-02-17340, 06-02-16702$ and $00-15-96814$), Russian state programs "Quantum Macrophysics", "Strongly Correlated electrons in Semiconductors, Metals, Superconductors and  Magnetic Materials", "Neutron Research of Solids"and Russian Belaruss cooperation  (grant $06-02-81029-Bel-a$).

\end{document}